\documentclass{elsart}
\usepackage{amssymb,amsmath,graphicx,bbm}
\begin{document}
\begin{frontmatter}
\title{Quantum communication with photon-number entangled states and
realistic photodetection}
\author{Vladyslav C. Usenko}
\address{Bogolyubov Institute for Theoretical Physics of 
the National Academy of Sciences, Kyiv, Ukraine \\
Department of Optics, Palack\' y University, Olomouc, Czech Republic}
\author{Matteo G. A. Paris}
\address{Dipartimento di Fisica, Universit\`a degli Studi di Milano, I-20133
Milano, Italy \\ CNISM, UdR Milano Universit\`a, I-20133 Milano, Italy\\ 
Institute for Scientific Interchange, I-10133 Torino, Italy}
\date{\today}
\begin{abstract}
We address the effects of realistic photodetection, with nonunit quantum
efficiency and background noise (dark counts), on the performances of
quantum communication schemes based on photon-number entangled states
(PNES).  We consider channels based on Gaussian twin-beam states (TWB) and
non-Gaussian two-mode coherent states (TMC) and evaluate the channel
capacity by optimizing the bit discrimination threshold. We found that
TWB-based channels are more robust against noise than TMC-based ones and
that this result is almost independent on the statistics of dark counts.
\end{abstract}
\maketitle
\end{frontmatter}
\section{Introduction}
Quantum communication schemes are aimed to improve security \cite{gisin}
and capacity \cite{drummond} of channels upon exploiting the specific
features of the involved quantum mechanical systems. Indeed,
quantum-enhanced key distribution (QKD) and communication schemes have been
developed for single qubit \cite{bb84} or entangled qubit pairs
\cite{e91}, and practically implemented using faint laser pulses or
photon pairs from spontaneous parametric downconversion. More recently,
much attention has been devoted to investigate the use of continuous
variable (CV) systems and protocols, in particular, using the sub-shot-noise fluctuations
of photon-number difference of two correlated beams \cite{funk}, the
sub-shot-noise modulations \cite{twb} and the sub-shot-noise
fluctuations of the photon numbers in each of the correlated modes
\cite{tmc2} have been proposed.  Although for these CV schemes the unconditional
security proofs have not been obtained yet \cite{gott}, they are of
interest and deserve investigations mostly due to the potential gain in
communication effectiveness.
\par
In this paper we consider schemes where information is encoded in the
degrees of freedom of a correlated state shared by the two parties. In
particular, we address binary communication channels based on
photon-number entangled states (PNES) \cite{bith} as the two-mode
coherently-correlated (TMC) or twin-beam  (TWB) states. The
communication protocol is based on photon number correlations and
realized upon choosing a shared threshold to convert the outcome of a
joint photon number measurement into a symbol from a discrete alphabet.
Notice that, in principle, entanglement itself is not needed to
establish this class of communication channels, which are based on
photon-number correlations owned also by separable mixed states.  On the
other hand, purity of the support state is relevant to increase security
of the channel, and the joint requirement of correlation and purity
leads to individuate PNES as a suitable choice for building effective
communication channels.  
\par
In this paper we consider PNES-based quantum communication with
realistic photodetection, which is affected both by nonunit quantum
efficiency and background noise (dark counts), and analyze the effects
of imperfections on the performances of channels. We consider channels
based on TWB and TMC and evaluate the channel capacity by optimizing the
bit discrimination threshold in the presence of noise. As we will see,
TWB-based channels are more robust against noise than TMC-based one and
this result is almost independent on the statistics of dark counts.
\par
The paper is structured as follows. At first we briefly 
introduce the two classes of PNES and illustrate the binary communication 
protocols. Then, we address noise in the decoding stage and derive 
the channel capacities in the presence of noise.
\section{Binary communication with photon number entangled states}
PNES are 
bipartite states of two modes of the field with Schmidt 
decomposition in the Fock number states. They may be written as 
\begin{equation}
\label{eq:twin}
\left| \Psi \right\rangle\rangle = \frac1{\sqrt{N}}
\sum\limits_n {\psi_n \left| {n,n} \right\rangle\rangle } ,
\end{equation}
where $\left| {n,n} \right\rangle\rangle = \left| 
n \right\rangle _1 \otimes \left| n 
\right\rangle _2$ and $N=\sum_n |\psi_n|^2$. 
PNES can be generated by means of parametric processes, either in
optical oscillators or amplifiers \cite{walls}. Generation of PNES have
been reported with photon number statistics varying from sub-Poisson
\cite{twb_exp} to super-Poisson \cite{twin1,twin2,twin3}.  As a matter
of fact, several quantum communication schemes and QKD protocols have
been proposed exploiting PNES correlations with coding based on the
beams intensity \cite{twb,tmc} or intensity difference \cite{zhang}. In
particular, the degenerate pair-coherent states
\cite{agarwal1,agarwal2}, also referred to as two-mode  coherently
correlated (TMC) \cite{tmc2} have been suggested as an effective
channel.  TMC may be written in the Fock basis as follows
\begin{equation}
\label{eq:TMC}
\left| \lambda \right\rangle\rangle = \frac{1}{\sqrt {I_0 \left( {2\left| \lambda 
\right|} \right)} }\sum\limits_n {\frac{\lambda ^n}{n!}\left| {n,n} 
\right\rangle\rangle } 
\end{equation}
The peculiarity of TMC is that they show sub-Poisson statistics for each 
of the beam. The corresponding QKD scheme is based on the fact that due to 
the strong intensity correlations one may decode a random bit sequence which will be 
correlated for the two remote sides carrying out independent but simultaneous 
intensity measurements on each of the two spatial twin modes. The realistic security 
of the scheme is based on the checking the beam statistics against 
the expected one corresponding to the fixed known state 
parameter. It was shown that realistic eavesdropping attempts 
cause statistics degradation to super-Poisson distribution and introduce 
perturbations in the obtained density matrix which are significant enough 
to be detected thus making eavesdropping ineffective \cite{tmc2}.
\par
The bits decoding for PNES-based communication protocol is quite
natural---each of the legitimate users measure the incoming photon 
number for a next time slot and compare the obtained value to a given 
bit threshold. If the current photon number 
value is above the threshold the corresponding bit value is considered to be 
equal to 1, while if the photon number is below the threshold the bit value 
is equal to 0:
\begin{eqnarray}
\label{eq:protocol}
B = \left\{ {{\begin{array}{*{20}c}
 { {n \le T}  \to 0} \hfill \\
 { {n > T}  \to 1} \hfill \\
\end{array} }} \right.
\end{eqnarray}
The threshold may be optimized, or set to a predetermined value, e.g. the 
integer part of the mean photon number.  With the latter choice the alphabet 
extension to the 4 and 8-letter sets was shown to increase the 
information capacity and make the protocol 
secure against intercept-resend attacks \cite{tmc}. 
\par
Another relevant class of the PNES states (\ref{eq:twin}) 
is the twin-beam state (TWB).  The Fock expansion is given by
\begin{equation}
\label{eq:twb}
\left| x \right\rangle\rangle = \sqrt {\left( {1 - x^2} \right)} \sum\limits_n 
{x^n\left| {n,n} \right\rangle\rangle }.
\end{equation}
TWB are Gaussian states \cite{twb2,twb3} and the photon statistics of 
the two modes is super-Poisson in contrast  to the sub-Poisson statistics 
of the TMC state modes. The TWB states can be used for the implementation 
of the same quantum communication 
protocol (\ref{eq:protocol}) exploiting the photon-number CV information coding, 
though the security issues remain an open question (especially the security 
against the intercept-resend eavesdropping, which is based on the sub-Poisson 
statistics check for the TMC). The TWB-based protocol may possibly require the 
use of the additional degrees of freedom in order to force the eavesdropper to 
guess the measurement and generation bases like it is in the celebrated BB84 
single-qubit QKD protocol \cite{bb84}.
\par
The practical implementation of PNES-based communication protocols
(\ref{eq:protocol}), either based on TMC or TWB, crucially depends on 
the influence of the realistic lossy optical media and realistic noisy
photodetectors. Those are the main sources of noise for the secure QKD
protocols implementation since they restrict the communication distances
and rates. The influence of losses on the performance of the PNES-based
channels (in comparison to classically mixed states) has been previously 
investigated \cite{bith}. Here we focus on the effect of the nonideal 
photodetection.
\section{Noisy photon-counting} In order to investigate the effect of the
noisy photon-counting on PNES-based quantum channel we model the
detector as an ideal one preceded by a beam-splitter of transmittivity
$\eta$ equal to the detection losses. The first port of the
beam-splitter is fed by the signal state, while the other port is
excited by an auxiliary state that reproduces the background noise.  The
action of the beam-splitter is described by the operator 
$U_{\phi}=\exp\{{\phi(a_1^{\dag}a_2-a_1a_2^{\dag})}\}$, 
which in the Heisenberg picture corresponds to the modes evolution \cite{gaussian}:
\begin{equation}
U_{\phi}^{\dag} \binom{a_1}{a_2}U_{\phi} = \mathbf{B}_{\phi}\binom{a_1}{a_2},
\mathbf{B}_{\phi} = \left ( \begin{array}{cc}
\cos{\phi} & \sin{\phi} \\
-\sin{\phi} & \cos{\phi}
\end{array} \right)\:,
\end{equation}
where $\eta=\cos{\phi}^2$.
The evolution of the Fock number basis may be expressed as 
\begin{equation}
U_{\phi}|n_1 \rangle \otimes |n_2 \rangle =\sum_{k_1=0}^{n_1}{\sum_{k_2=0}^{n_2}{A_{k_1k_2}^{n_1 n_2}|k_1+k_2
\rangle \otimes |n_1+n_2-k_1-k_2 \rangle}}
\end{equation}
where the transfer matrix $A_{k_1k_2}^{n_1 n_2}$ is given by
\begin{eqnarray}
\lefteqn{A_{k_1k_2}^{n_1n_2}=\sqrt{
\frac{(k_1+k_2)!(n_1+n_2-k_1-k_2)!}{n_1!n_2!}}(-1)^{k_2}\times{}}
\nonumber \\
{}&&\times\binom{n_1}{k_1} \binom{n_2}{k_2}
\sin{\phi}^{n_1-k_1+k_2}
\cos{\phi}^{n_2+k_1-k_2}
\label{eq:A}
\end{eqnarray}
Upon writing the signal as $\rho=\sum_{nm}{\rho_{nm}|n\rangle\langle m|}$ 
and the noise state as $\nu=\sum_{p}\nu_p{|p\rangle \langle p|}$, then the 
probability to have $s$ {\em counts} at the output is given by 
\begin{equation}
p_s=Tr[U_{\phi}\,\rho\otimes\nu\, U_{\phi}^{\dag} |s\rangle \langle s| 
\otimes \mathbb{I}]\:,
\label{counts}
\end{equation}
which may be written as 
\begin{align}
p_s&=
\sum_{n=0}^\infty \rho_{nn} \left[
\sum_{p=0}^\infty \nu_p \left(\sum_{k=0}^s A_{k,s-k}^{n,p}\right)^2 
\theta(n+p-s)\right] = \nonumber \\
&= \frac{1-\eta}{\eta}^s\sum_{n=0}^{\infty}{ \sum_{p=0}^{\infty}{
(1-\eta)^n \, \rho_{nn} \, \eta^p \, \nu_p
\, \binom{n+p-s}{p-s}
\, \binom{p}{s}}}\times{} \nonumber\\
&  \times {}_2F_1({-n,-s,1+p-s,-\frac{\eta}{1-\eta}})^2
\: \theta(n+p-s)\,,
\label{pom}
\end{align}
$\theta(x)$ being the Heaviside step function.
Eqs. (\ref{counts}) and (\ref{pom}) express the relation between the probability of 
photocounts in a noisy detector and the actual photon distribution of
the incoming signal.
Once the joint probability distribution is known we may evaluate
the mutual information between the two parties and optimize it 
against the threshold. For the binary coding of Eq. (\ref{eq:protocol}) 
the two parties infer the same symbol with probabilities
\begin{align}
p_{00} = \sum_{p=0}^T \sum_{q=0}^T P_\eta (p,q) \quad 
p_{11} = \sum_{p=T}^\infty \sum_{q=T}^\infty P_\eta (p,q)
\label{p00p11}\;.
\end{align}
In the ideal case, {\em i.e.} with no losses, PNES-based 
protocols achieve $p_{00}+p_{11}=1$, due to perfect correlations
between the two modes. On the other hand, if $\eta\neq 1$ the unwanted 
inference events "01" and "10" may occur with probabilities
\begin{align}
p_{01} = \sum_{p=0}^T \sum_{q=T}^\infty P_\eta (p,q) \quad 
p_{10} = \sum_{p=T}^\infty \sum_{q=0}^T P_\eta (p,q)
\label{p10p01}\;.
\end{align}
The probabilities are not independent since the normalization 
condition $p_{00}+p_{10}+p_{01}+p_{11}=1$ holds.  
The mutual information between the two alphabets reads as follows
\begin{align}
I_2 = \sum_{i=0}^1 \sum_{j=0}^1 p_{ij} \log \frac{p_{ij}}{q_i r_j}
\label{IM}\;,
\end{align}
where 
\begin{align}
q_i=p_{i0}+p_{i1} \quad i=0,1 \quad r_j=p_{0j}+p_{1j} \quad j=0,1 \:,
\end{align}
represents the marginal probabilities, {\em i.e.} the unconditional
probabilities of inferring the symbol ``i'' (``j'') for the first
(second) party. The mutual information, once the average number of input
photons and the loss parameter have been set, depends only on the
threshold value $T$. The channel capacity ${\cal C}=\max_T I_2$
corresponds to the maximum of the mutual information over the threshold.
We have obtained the channel capacity numerically by looking for the
optimal bit discrimination threshold as a function of the input energy
and of the intensity of the noise states, assuming that detectors are
equivalent for the both modes. In our calculations we considered 
the background noise either with Poisson statistics so that
$\nu_p=e^{-N}\frac{N^p}{p!}$, or thermal statistics 
$\nu_p=\frac{N^p}{(N+1)^{p+1}}$, where $N$
is the average number of photons of a noise mode.
\par
\begin{figure}
\includegraphics[width=0.41\textwidth]{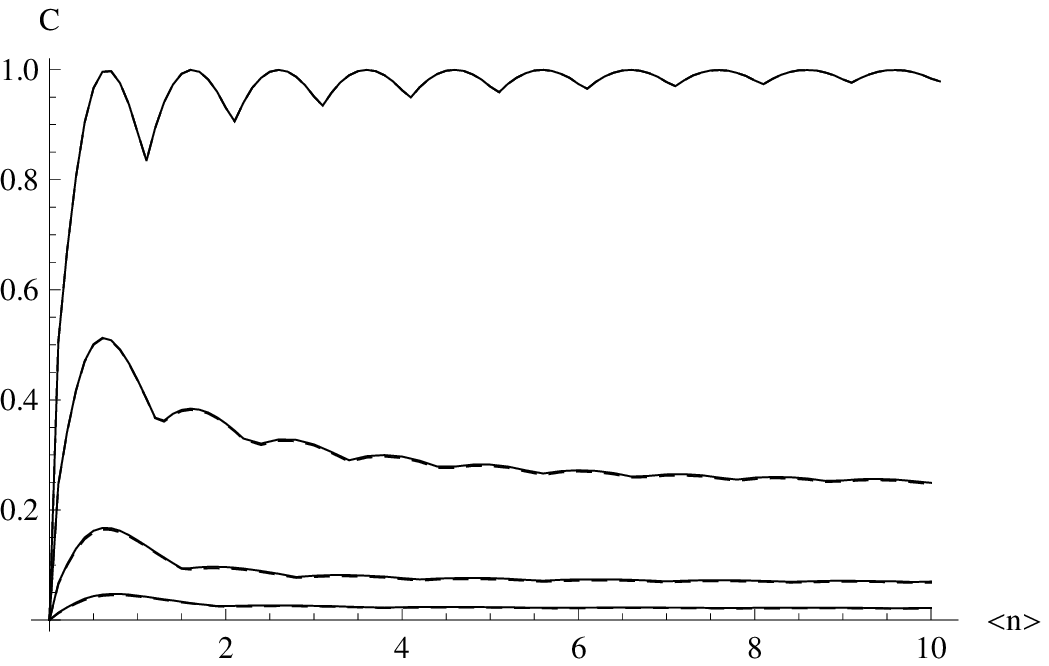} 
$\quad$
\includegraphics[width=0.41\textwidth]{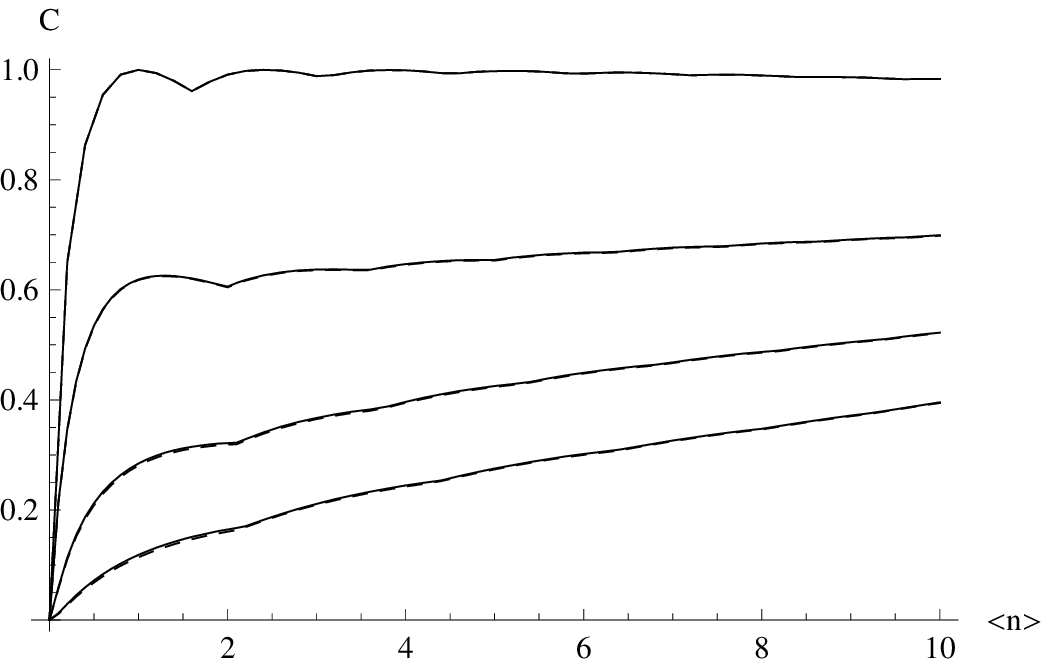} 
\caption{Channel capacity (mutual information maximized over the bit
threshold) for the TMC- (left) and TWB-based quantum channels as a
function of their average number of photons for different values of the
detection efficiency $\eta$; from bottom to top: $\eta=0.5, \eta=0.7,
\eta=0.9$ and $\eta=1$, the latter corresponding to the ideal detectors.
The detection noise is $N=0.2$, noise statistics is either thermal
(solid lines) or Poisson (dashed lines), almost coinciding.}
\label{PNES-CAP}
\end{figure}
\begin{figure}
\includegraphics[width=0.41\textwidth]{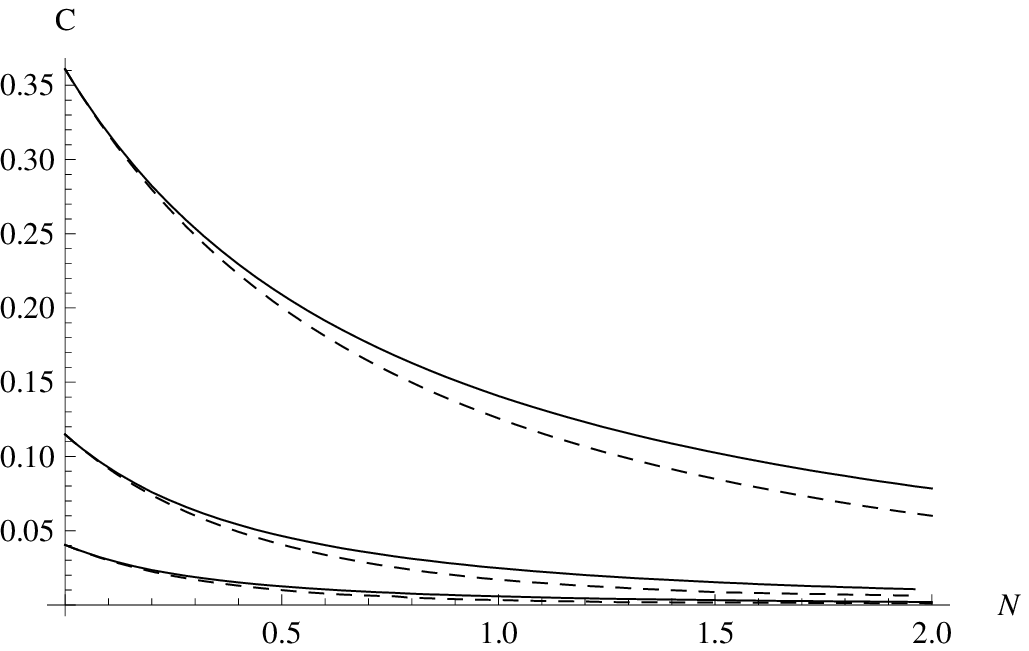} 
$\quad$
\includegraphics[width=0.41\textwidth]{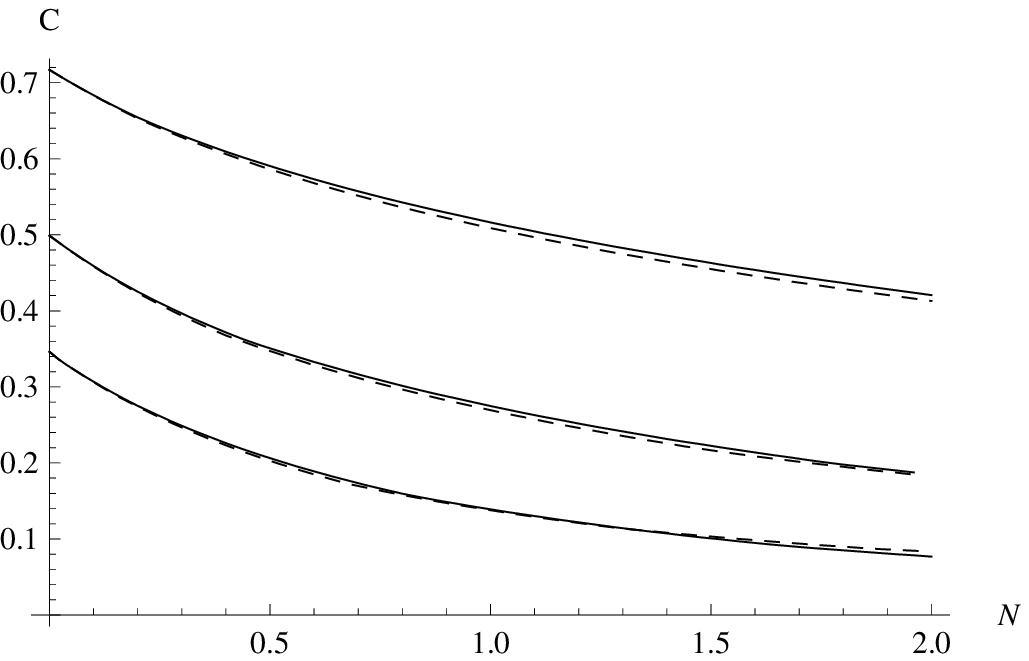} 
\caption{Channel capacity (mutual information maximized over the bit
threshold) for the TMC- (left) and TWB-based quantum channels as a
function of the detector noise average number of photons for different
values of the loss parameter $\eta$; from bottom to top: $\eta=0.5,
\eta=0.7, \eta=0.9$, average number of photons in a signal mode is
$\langle n \rangle = 5$, noise statistics is thermal (solid lines) or
Poisson (dashed lines).} \label{PNES-CAP2} \end{figure}

The channel capacities for both TMC and TWB states versus a signal mode
intensity are shown in Fig. \ref{PNES-CAP} for noise average photon
number $N=0.2$ and various detector efficiences $\eta$. At fixed energy
the channel capacity is larger for TWB than for TMC, even when being
reduced by an inefficient detection. The channel capacities were also
calculated for the fixed signal energy with respect to increasing noise
intensity. The results are given at Fig. \ref{PNES-CAP2}, the signal
mode average photon number is $\langle n \rangle = 5$. Again, the TWB
states are mode effective than TMC at the same energy and appear to be
more robust against the detection noise. Also, the results are almost
indistinguishable for Poisson and thermal noise, the latter being a 
bit more destructive, especially for TMC states for higher noise 
intensities.
\section{Conclusions}
We have analyzed the effect of detection noise,
quantum efficiency and dark cnounts, on the performance of PNES-based 
quantum channels. Our results show that the TWB-based channels are more 
robust against noise. The statistical properties of noise do not play 
a significant role, as the noise impact mostly depends on its intensity.
\section*{Acknowledgements}
V.U. acknowledges the support from Landau Network-Centro Volta through the 
Cariplo Foundation fellowship and from the project No. 202/07/J040 of Grant Agency of Czech Republic.

\end{document}